\documentclass[structabstract]{aa}
\usepackage{natbib}
\usepackage{graphicx}
\usepackage{epstopdf}
\usepackage{longtable}
\usepackage{lscape}
\begin{document}
   \title{The FERRUM project: laboratory-measured transition probabilities for \ion{Cr}{ii}}
\authorrunning{Gurell et al.}
\titlerunning{The FERRUM project: laboratory-measured transition probabilities for \ion{Cr}{ii}}

   \author{J. Gurell
          \inst{1}
          \and
          H. Nilsson\inst{2}
          \and
          L. Engstr\"om\inst{3}
          \and
          H. Lundberg\inst{3}
          \and
          R. Blackwell-Whitehead\inst{2}
          \and
          \\K. E. Nielsen\inst{4,5}
          \and
          S. Mannervik\inst{1}
          }

   \institute{Department of Physics, Stockholm University, AlbaNova University Center, SE-10691 Stockholm, Sweden\\
              \email{jonas.gurell@fysik.su.se}
         \and
             Lund Observatory, Lund University, Box 43, SE-22100 Lund, Sweden
             \and
                Atomic Physics, Department of Physics, Lund University, Box 118, SE-22100 Lund, Sweden
                \and
            Catholic University of America, Washington, DC 20064, USA
            \and
            Astrophysics Science Division, Code 667, Goddard Space Flight Center, Greenbelt, MD 20771, USA
             }

   \date{Received XXX; accepted XXX}


  \abstract
   {}
   {We measure transition probabilities for \ion{Cr}{ii} transitions from the $z$\,$^4$H$_J$, $z$\,$^2$D$_J$, $y$\,$^4$F$_J$, and $y$\,$^4$G$_J$ levels in the energy
   range 63000 to 68000\,cm$^{-1}$.}
   {Radiative lifetimes were measured using time-resolved laser-induced fluorescence from a laser-produced plasma. In addition, branching fractions were
   determined from intensity-calibrated spectra recorded with a UV Fourier transform spectrometer. The branching fractions and radiative lifetimes were combined
   to yield accurate transition probabilities and oscillator strengths.}
   {We present laboratory measured transition probabilities for 145 \ion{Cr}{ii} lines and radiative lifetimes for 14 \ion{Cr}{ii} levels. The laboratory-measured
   transition probabilities are compared to the values from semi-empirical calculations and laboratory measurements in the literature.}
   {}


   \maketitle

\section{Introduction}
Spectral analysis of astrophysical objects depends on the availability of accurate laboratory data including radiative lifetimes and transition probabilities.
Lines of \ion{Cr}{ii} are observed in a broad range of stellar and nebular spectra (e.g. \citeauthor{Meril}~\citeyear{Meril};
\citeauthor{Shevchenko}~\citeyear{Shevchenko}; \citeauthor{Andrievsky}~\citeyear{Andrievsky}), and accurate Cr data are required for stellar abundance studies
(\citeauthor{Babel}~\citeyear{Babel}; \citeauthor{Dimitrijevic}~\citeyear{Dimitrijevic}). In particular, several chemically peculiar stars show unexpectedly
high abundances of Cr (\citeauthor{Rice}~\citeyear{Rice}; \citeauthor{Lopez}~\citeyear{Lopez}).

Radiative lifetimes in \ion{Cr}{ii} have been with the beam-foil technique by \citeauthor{Pinnington1973}~(\citeyear{Pinnington1973}) and
\citeauthor{Engman}~(\citeyear{Engman}) and with the time-resolved laser-induced fluorescence (TRLIF) technique by \citeauthor{Schade}~(\citeyear{Schade}),
\citeauthor{Pinnington1993}~(\citeyear{Pinnington1993}), and \citeauthor{Nilsson}~(\citeyear{Nilsson}). In addition, branching fraction ($BF$) measurements
were combined with radiative lifetimes to yield transition probabilities (\citeauthor{bergeson}~\citeyear{bergeson}; \citeauthor{Spreger}~\citeyear{Spreger};
\citeauthor{Gonzalez}~\citeyear{Gonzalez}; \citeauthor{Nilsson}~\citeyear{Nilsson}), and oscillator strengths were measured by
\citeauthor{Musielok}~(\citeyear{Musielok}), \citeauthor{Goly}~(\citeyear{Goly}) and \citeauthor{Wujec}~(\citeyear{Wujec}) using a wall-stabilized arc.
Semi-empirical oscillator strengths have been calculated by \citeauthor{Kurucz2}~(\citeyear{Kurucz2}) using the Cowan code, by
\citeauthor{Luke}~(\citeyear{Luke}) using the R matrix method, and by \citeauthor{RU}~(\citeyear{RU}) using the orthogonal operator method.

Several studies have found that the stellar chromium abundance determined from \ion{Cr}{i} lines is significantly different to the abundance using \ion{Cr}{ii}
lines (\citeauthor{McWilliam}\,\citeyear{McWilliam};\citeauthor{Sobeck}\,\citeyear{Sobeck};\citeauthor{Lai}\,\citeyear{Lai}). This difference is greater than
the uncertainty in the stellar observations and measured oscillator strengths. Furthermore, the difference increases as the metallicity of the star decreases.
\citeauthor{Sobeck} (\citeyear{Sobeck}) indicate that one possible explanation of this discrepancy could be non-LTE effects not included in the stellar model.
\citeauthor{Sobeck} (\citeyear{Sobeck}) propose that, in order to resolve this issue additional laboratory investigations of chromium should focus on weak
branches of \ion{Cr}{ii} and that the Cr abundance should be reanalyzed with a three-dimensional hydrodynamical model. In addition, the need for more
laboratory measured \ion{Cr}{ii} transition probabilities is discussed by \citeauthor{Wallace} (\citeyear{Wallace}).


   \begin{figure*}
   \centering
  \includegraphics{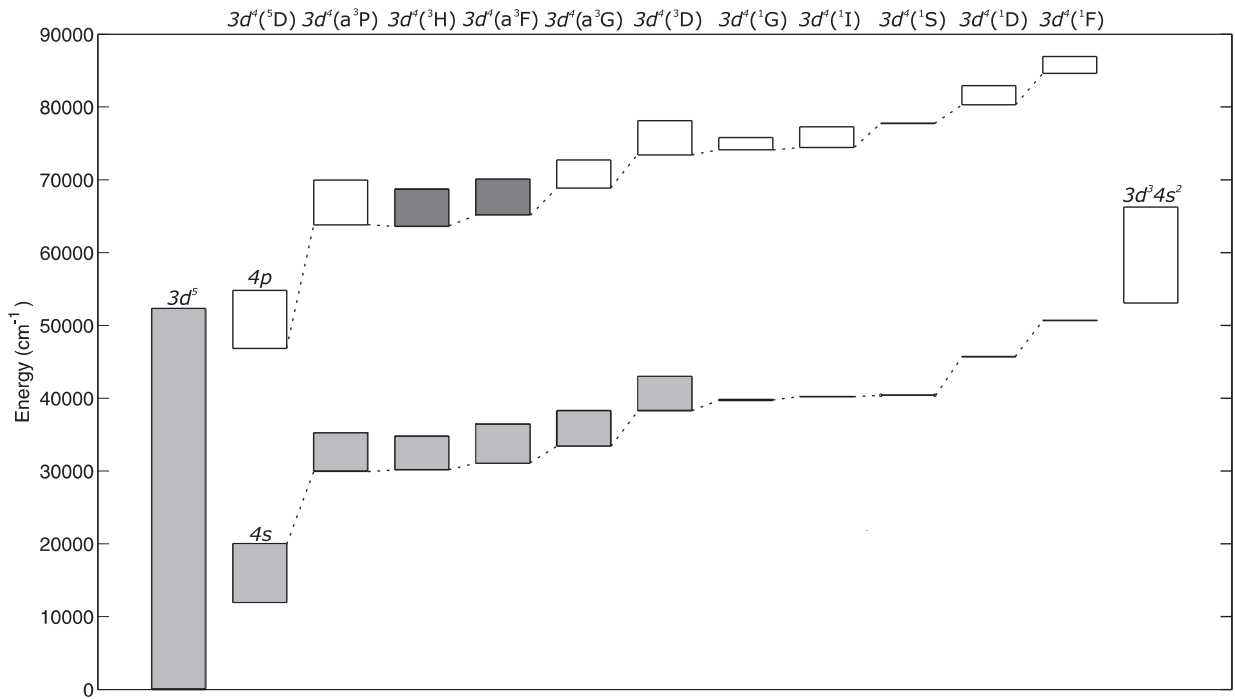}
   \caption{Partial energy level diagram of \ion{Cr}{ii} (\citeauthor{NIST}~\citeyear{NIST}). Only terms belonging to the $3d^5$, $3d^44s$, $3d^44p$, and $3d^34s^2$
   configurations are shown and ordered by their respective parent term shown above the diagram. The investigated upper levels are shown as dark grey boxes and the
   lower levels to which they decay are shown as light grey boxes.} \label{CrII}
              \label{FigGam}%
    \end{figure*}

In this paper we present transition probabilities for 145 lines in \ion{Cr}{ii} from 14 upper levels, see the partial energy level diagram of \ion{Cr}{ii} in
Figure~\ref{CrII}. The lifetimes of the upper levels have been measured with the TRLIF technique.

\section{Laboratory measurements}

The lifetime of an upper state $i$ can be written as

\begin{equation}
\tau_{\rm i}=1/\sum\limits_{k}A_{{\rm ik}},
\end{equation}
where $A_{ik}$ is the transition probability of a line from upper level $i$ to lower level $k$. The $BF$ of the line is defined as

\begin{equation}
BF_{ik}=A_{ik}/\sum\limits_{k}A_{ ik}=I_{ik}/\sum\limits_{k}I_{ ik}, \label{bfs}
\end{equation}
where $I_{ik}$ is the measured intensity corrected for the instrumental response. Combining these two equations gives the transition probabilities as
\begin{equation}
A_{ik}=BF_{ik}/\tau_{\rm i}. \label{A}
\end{equation}
The following subsections describe the measurements of the lifetimes and $BF$s.

\subsection{Radiative lifetimes}

We measured lifetimes for 14 odd parity levels in \ion{Cr}{ii} belonging to the $3d^44p$ configuration. The lifetimes were measured using the TRLIF technique
at the Lund High Power Laser Facility. This technique has been described in detail in the literature (see \citeauthor{Bergstrom1988}~\citeyear{Bergstrom1988},
\citeauthor{Xu2003}~\citeyear{Xu2003}), so only a brief description is given here.

A laser-produced `plasma-cone' containing Cr atoms and ions in metastable levels was created by focusing a Nd:YAG laser (Continuum Surelite) onto a target of
pure Cr. Cr$^+$ ions in metastable states were excited to levels of opposite parity using a pump laser and the fluorescence from the excited levels was
recorded as a function of time. The excitation pulses were created by pumping a Continuum Nd-60 dye laser with a Nd:YAG Continuum NY-82 laser. The pulses from
the Nd:YAG pump laser were shortened from 10 to 1.5 ns using stimulated Brillouin scattering. The Nd-60 dye laser used a DCM dye to produce light between 6000
and 6700 \AA. A broader wavelength coverage was achieved by using nonlinear effects in KDP and BBO crystals and Raman shifts in a H$_2$ cell.

A 1/8 m monochromator was used to select the observable fluorescence wavelength. The fluorescence signal was recorded with a micro-channel plate
photomultiplier tube with a rise time of 0.2 ns. The shape of the excitation pulse was measured with the same system as the fluorescence signal. The lifetimes
were extracted by fitting the fluorescence data with a single exponential convoluted with the shape of the laser pulse. Each lifetime curve was averaged over
1000 laser shots, and the final lifetimes given in Table~\ref{Lifetimes} are averages of at least 10 lifetime curves. The uncertainties in the lifetimes
include both statistical and systematic errors.

\subsection{Branching fractions}

\begin{figure}
   \centering
   \includegraphics[width=8cm]{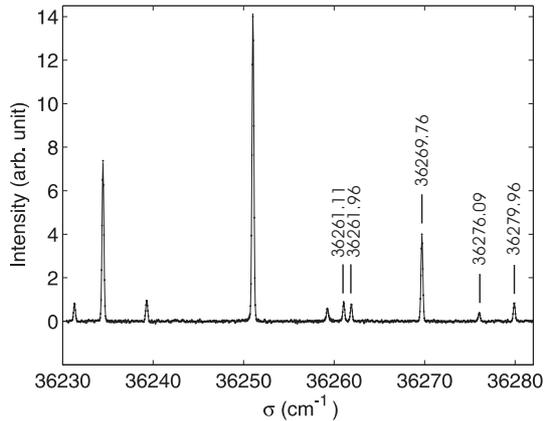}
      \caption{A section of the FT spectrum showing five \ion{Cr}{ii} lines indicated by their respective wavenumber.} \label{spectrum}
   \end{figure}

The $BF$s were measured from spectra recorded with the Chelsea Instrument FT500 UV FT spectrometer at Lund Observatory. The light source was a Penning
discharge lamp with pure Cr cathodes and operated with Ne as buffer gas. The light source was operated at a current between 0.8 and 1.3 A, and with at a
carrier gas pressure of  40 mTorr. The Penning discharge lamp provides an intensity stable emission spectrum over several hours enabling high signal-to-noise
(S$/$N) spectra to be recorded. Figure~\ref{spectrum} shows part of the observed spectrum.

Three separate spectral regions were recorded to cover the wavenumber region between 20000 and 52000\,cm$^{-1}$. The spectra were intensity-calibrated with
standard lamps with known spectral radiances and by \ion{Cr}{ii} $BF$s previously measured by \citeauthor{Nilsson}~(\citeyear{Nilsson}). A continuous deuterium
(D$_2$) lamp was used  in the wavenumber range $32000-52000$\,cm$^{-1}$. A tungsten strip lamp was used in the wavenumber range $20000-28000$\,cm$^{-1}$ and
the 28000-32000\,cm$^{-1}$ wavenumber region was calibrated using previously intensity-calibrated \ion{Cr}{ii} $BF$s from the data of
\citeauthor{Nilsson}~(\citeyear{Nilsson}). The calibration was performed with a software routine implemented in the program XGremlin (Nave et al. 1997). In
addition, the spectra were wavenumber-calibrated using an average of nine unblended \ion{Ne}{ii} lines, which have been measured with high accuracy and
suggested as suitable transitions for wavenumber calibration by \citeauthor{Oberg}~(\citeyear{Oberg}).

However, the wavelengths and wavenumbers in Tables~\ref{BFs} and \ref{Linelist} are Ritz values determined from the energy levels in
\citeauthor{NIST}~(\citeyear{NIST}).

The spectral lines were fitted with Voigt profiles to determine the integrated intensity using the commercially available software PeakFit. The uncertainty in
the integrated intensity is determined from the standard deviation in the fitted values and the S$/$N of the lines. The majority of the fitted \ion{Cr}{ii}
line profiles were unblended. However, a small number of \ion{Cr}{ii} lines were partially blended with other lines. The blended features were identified, and
a fit of both line profiles was performed when possible. The blended \ion{Cr}{ii} lines are given with larger uncertainties in the integrated intensity.

\section{Results}

Our experimental lifetimes are compared to previous measurements and calculations in Table~\ref{Lifetimes}. Table~\ref{BFs} shows our laboratory measured
$BF$s, $A$-values and semi-empirical $BF$s. The complete line list for all \ion{Cr}{ii} lines measured in this paper with comparisons to semi-empirical
log~$gf$ values in the literature is shown in Table~\ref{Linelist}. A comparison between our measured $A$-values and values presented in the literature is
shown in Table~\ref{Acomp}.

\begin{table*}
\caption{Lifetimes in \ion{Cr}{ii}} \label{Lifetimes} \centering
\begin{tabular}{@{}llccccccccc}     
\hline\hline
& & & \multicolumn{8}{c}{$\tau$ (ns)} \\
\cline{4-11}
Configuration & Term &  $E$ (cm$^{-1}$)  & This work & C\&B$^a$ & WT$^b$ & P$^c$ & WE$^d$ & E$^e$ & L$^f$ & R\&U$^g$ \\
\hline
 ($a$\,$^3$H)$4p$ & $z$\,$^4$H$^o_{7/2}$ & 63600 & 4.4(4) & 2.1(4)$^h$ & 3.3$^h$ & 4.0(4)$^h$ & 5.3$^{h,i}$ & &3.3 & 4.4\\
 ($a$\,$^3$H)$4p$ & $z$\,$^4$H$^o_{9/2}$ & 63706 & 4.4(4) & 2.1(4)$^h$ & 3.3$^h$ & 4.0(4)$^h$ & 5.3$^{h,i}$ & &3.3 & 4.4\\
 ($a$\,$^3$H)$4p$ & $z$\,$^4$H$^o_{11/2}$ & 63849 & 4.2(3) & 2.1(4)$^h$ & 3.3$^h$ & 4.0(4)$^h$ & 5.3$^{h,i}$ & 5.0(5) &3.3 & 4.4\\
 ($a$\,$^3$H)$4p$ & $z$\,$^4$H$^o_{13/2}$ & 64031 & 4.2(3) & 2.1(4)$^h$ & 3.3$^h$ & 4.0(4)$^h$ & 5.3$^{h,i}$ & 4.7(5) &3.3 & 4.3\\
 & & & & & & & & & & \\
 ($a$\,$^3$F)$4p$ & $y$\,$^4$F$^o_{5/2}$ & 67012 & 3.7(4) & & &  & & & & 4.3\\
 ($a$\,$^3$F)$4p$ & $y$\,$^4$F$^o_{3/2}$ & 67070 & 3.9(5) & & &  & & & & 4.5\\
 ($a$\,$^3$F)$4p$ & $y$\,$^4$F$^o_{7/2}$ & 67393 & 2.9(2) & & &  & & & & 2.9\\
 ($a$\,$^3$F)$4p$ & $y$\,$^4$F$^o_{9/2}$ & 67448 & 2.9(2) & & & & & & & 2.9\\
 & & & & & & & & & &\\
 ($a$\,$^3$H)$4p$ & $y$\,$^4$G$^o_{7/2}$ & 67333 & 2.6(2) & & &  3.4(2)$^h$ & & & & 2.6\\
 ($a$\,$^3$H)$4p$ & $y$\,$^4$G$^o_{5/2}$ & 67344 & 2.6(2) & & &  3.4(2)$^h$ & & & & 2.5 \\
 ($a$\,$^3$H)$4p$ & $y$\,$^4$G$^o_{9/2}$ & 67353 & 2.6(2) & & &  3.4(2)$^h$ & & & & 2.5\\
 ($a$\,$^3$H)$4p$ & $y$\,$^4$G$^o_{11/2}$ & 67369 & 2.7(2) & & &  3.4(2)$^h$ & & & & 2.5\\
 & & & & & & & & & & \\
 ($a$\,$^3$F)$4p$ & $z$\,$^2$D$^o_{3/2}$ & 67379 & 3.1(3) & & &  & & & & 3.3\\
 ($a$\,$^3$F)$4p$ & $z$\,$^2$D$^o_{5/2}$ & 67387 & 3.1(3) & & &  & & & & 3.4\\
\hline
 \multicolumn{6}{l}{$^a$Corliss and Bozman (1962) (Intensity measurement)}\\
 \multicolumn{7}{l}{$^b$Warner (1967) (Calculated)}\\
 \multicolumn{7}{l}{$^c$Pinnington et al.(1973) (Beam-foil measurement)}\\
\multicolumn{7}{l}{$^d$Warner (1967) (Emission measurement)}\\
\multicolumn{7}{l}{$^e$Engman (1975) (Beam-foil measurement)}\\
 \multicolumn{7}{l}{$^f$Luke (1988) (Calculated)}\\
 \multicolumn{7}{l}{$^g$\citeauthor{RU} (\citeyear{RU}) (Calculated)}\\
\multicolumn{8}{l}{$^h$Mean lifetime}\\
\multicolumn{8}{l}{$^i$Upper limit}\\
\end{tabular}
\end{table*}

\subsection{Lifetimes}

In Table~\ref{Lifetimes} we report radiative lifetimes for 14 levels in \ion{Cr}{ii} compared with values from the literature. Our experimental lifetimes of
the $z$\,$^4$H term agree, within the uncertainties, with the beam-foil measurements by \citeauthor{Pinnington1973}~(\citeyear{Pinnington1973}) and by
\citeauthor{Engman}~(\citeyear{Engman}), and the experimental lifetimes of \citeauthor{Warner}~(\citeyear{Warner}). There is a good agreement between our
lifetimes and the values from \citeauthor{RU}~(\citeyear{RU}); however, the calculated values by \citeauthor{Warner}~(\citeyear{Warner}) and
\citeauthor{Luke}~(\citeyear{Luke}) are shorter than our lifetimes by approximately four standard deviations.

There is a a three-sigma difference between our lifetimes for the $y$\,$^4$G term and the measured values of
\citeauthor{Pinnington1973}~(\citeyear{Pinnington1973}). The difference may come from the improved wavelength resolution of our measurements.
\citeauthor{Pinnington1973}~(\citeyear{Pinnington1973}) were not able to resolve the four individual levels $y$\,$^4$G and measured an average lifetime
determined from a blend of transitions from these levels. \citeauthor{Pinnington1973}~(\citeyear{Pinnington1973}) also states that the assignment of the
transition from the $y$\,$^4$G term at 2700~\AA~is one of the `less certain' ones presented. In our measurements we have resolved the individual decay
channels.

There is good agreement between our work and the lifetimes calculated by \citeauthor{RU}~(\citeyear{RU}) for the $y\,^4$F$_{9/2, \,7/2}$ levels. However, there is a
discrepancy between our lifetimes and the values of \citeauthor{RU}~(\citeyear{RU}) for $y\,^4$F$_{5/2, \,3/2}$ and the $z\,^2$D$_{5/2, \,3/2}$. This may be
due to a difference in the predicted level mixing, which is difficult to reproduce in semi-empirical calculations for complex atomic systems such as
\ion{Cr}{ii}. In addition, an increase in the predicted level mixing may explain the discrepancy between our experimental $BF$s and the calculated $BF$s by
\citeauthor{RU}~(\citeyear{RU}) for transitions from the $y\,^4$F$_{5/2, \,3/2}$ and the $z\,^2$D$_{5/2, \,3/2}$ levels.

\subsection{Transition probabilities}
In Table~\ref{BFs} we present our $BF$s and $A$-values. The $BF$ residual value in Table~\ref{BFs} is determined from the semi-empirical calculations of
\citeauthor{RU} (\citeyear{RU}). The residual value is used to estimate the $BF$ contribution from transitions that were not observed in our spectra. The
missing lines are very weak transitions, $BF\leq 2\%$, and for most upper levels in Table~\ref{BFs}, the residual value is at most a few percent, which is less
than the uncertainty in the $BF$s. The uncertainty in the $A$-values is determined from the $BF$ and lifetime uncertainty using the method discussed by
\citeauthor{Sikstrom}~(\citeyear{Sikstrom}), which include uncertainties from the line-fitting, intensity calibration of each spectra, intensity cross
calibration between separate spectra and the uncertainty in the fit of the decay curve.



\onllongtab{2}{
\begin{longtable}{@{}llcccccc}
\caption{\label{BFs} Observed transitions, experimental $BF$s and $A$-values presented along with calculated $BF$s using the orthogonal operator approach (\citeauthor{RU}~\citeyear{RU})}\\
\hline\hline
Upper & Lower &  $\sigma$ & $\lambda_{air}$ & \multicolumn{2}{c}{$BF$}  & $A$ & Unc\\
\cline{5-6}
level & level &  (cm$^{-1}$)  &  (\AA) & Exp. & Theory $^a$& ($10^7$\,s$^{-1}$) &(\%)\\
\hline
\endfirsthead
\caption{continued.}\\
\hline\hline
Upper & Lower &  $\sigma$ & $\lambda_{air}$ & \multicolumn{2}{c}{$BF$}& $A$ & Unc\\
\cline{5-6}
level & level &  (cm$^{-1}$)  &  (\AA) & Exp. & Theory$^a$ & ($10^7$\,s$^{-1}$) &(\%)\\
\hline
\endhead
\hline
\endfoot
$3d^4$($^3$H)$4p$   $^4$H$_{13/2}^o$    &   $3d^4$($^3$H)$4s$   $^2$H$_{11/2}$  &   29217.60    &   3421.613    &   0.002   &   0.004   &   0.06    &   20  \\
        &   $3d^4$($^3$G)$4s$   $^4$G$_{11/2}$  &   30336.36    &   3295.425    &   0.042   &   0.089   &   0.99    &   14  \\
        &   $3d^4$($^3$H)$4s$   $^4$H$_{13/2}$  &   33638.67    &   2971.902    &   0.819   &   0.771   &   19.50   &   7   \\
        &   $3d^5$  $^2$I$_{13/2}$  &   33880.64    &   2950.679    &   0.007   &   0.001   &   0.16    &   32  \\
        &   $3d^5$  $^4$G$_{11/2}$  &   43518.41    &   2297.171    &   0.125   &   0.135   &   2.97    &   14  \\
        &\textit{Residual}          &       &       &   0.002   &       &       &       \\
        &           &       &       &       &       &       &       \\
$3d^4$($^3$H)$4p$   $^4$H$_{11/2}^o$    &   $3d^4$($^3$H)$4s$   $^2$H$_{9/2}$   &   29217.80    &   3421.591    &   0.002   &   0.001   &   0.04    &   76  \\
        &   $3d^4$($^3$G)$4s$   $^4$G$_{11/2}$  &   30154.56    &   3315.294    &   0.002   &   0.005   &   0.05    &   14  \\
        &   $3d^4$($^3$G)$4s$   $^4$G$_{9/2}$   &   30229.75    &   3307.047    &   0.041   &   0.083   &   0.99    &   14  \\
        &   $3d^5$  $^4$F$_{9/2}$   &   30994.42    &   3225.456    &   0.002   &   0.004   &   0.04    &   16  \\
        &   $3d^4$($^3$F)$4s$   $^4$F$_{9/2}$   &   32629.36    &   3063.834    &   0.011   &   0.008   &   0.27    &   14  \\
        &   $3d^4$($^3$H)$4s$   $^4$H$_{13/2}$  &   33456.86    &   2988.052    &   0.038   &   0.034   &   0.90    &   14  \\
        &   $3d^4$($^3$H)$4s$   $^4$H$_{11/2}$  &   33550.22    &   2979.737    &   0.780   &   0.736   &   18.56   &   7   \\
        &   $3d^5$  $^4$G$_{9/2}$   &   43329.42    &   2307.191    &   0.112   &   0.120   &   2.66    &   14  \\
        &   $3d^5$  $^4$G$_{11/2}$  &   43336.59    &   2306.809    &   0.009   &   0.009   &   0.21    &   16  \\
        &\textit{Residual}          &       &       &   0.004   &       &       &       \\
        &           &       &       &       &       &       &       \\
$3d^4$($^3$H)$4p$   $^4$H$_{9/2}^o$ &   $3d^4$($^3$G)$4s$   $^4$G$_{9/2}$   &   30087.32    &   3322.704    &   0.004   &   0.008   &   0.09    &   18  \\
        &   $3d^4$($^3$G)$4s$   $^4$G$_{7/2}$   &   30185.16    &   3311.933    &   0.042   &   0.083   &   0.95    &   15  \\
        &   $3d^5$  $^4$F$_{7/2}$   &   30869.55    &   3238.504    &   0.003   &   0.005   &   0.07    &   20  \\
        &   $3d^4$($^3$F)$4s$   $^4$F$_{7/2}$   &   32537.67    &   3072.469    &   0.013   &   0.010   &   0.29    &   17  \\
        &   $3d^4$($^3$H)$4s$   $^4$H$_{11/2}$  &   33407.78    &   2992.442    &   0.057   &   0.054   &   1.29    &   15  \\
        &   $3d^4$($^3$H)$4s$   $^4$H$_{9/2}$   &   33487.47    &   2985.321    &   0.761   &   0.704   &   17.29   &   9   \\
        &   $3d^5$  $^4$G$_{9/2}$   &   43186.97    &   2314.802    &   0.013   &   0.013   &   0.29    &   16  \\
        &   $3d^5$  $^4$G$_{7/2}$   &   43188.45    &   2314.722    &   0.101   &   0.116   &   2.28    &   15  \\
        &\textit{Residual}          &       &       &   0.008   &       &       &       \\
        &           &       &       &       &       &       &       \\
$3d^4$($^3$H)$4p$   $^4$H$_{7/2}^o$ &   $3d^5$  $^2$G$_{9/2}$   &   27328.36    &   3658.160    &   0.014   &   0.000   &   0.33    &   15  \\
        &   $3d^4$($^3$G)$4s$   $^4$G$_{7/2}$   &   30079.75    &   3323.539    &   0.003   &   0.008   &   0.08    &   22  \\
        &   $3d^4$($^3$G)$4s$   $^4$G$_{5/2}$   &   30182.87    &   3312.184    &   0.045   &   0.088   &   1.02    &   15  \\
        &   $3d^4$($^3$F)$4s$   $^4$F$_{5/2}$   &   32483.54    &   3077.588    &   0.009   &   0.005   &   0.20    &   20  \\
        &   $3d^4$($^3$H)$4s$   $^4$H$_{9/2}$   &   33382.07    &   2994.747    &   0.056   &   0.051   &   1.27    &   15  \\
        &   $3d^4$($^3$H)$4s$   $^4$H$_{7/2}$   &   33444.12    &   2989.190    &   0.766   &   0.709   &   17.40   &   9   \\
        &   $3d^5$  $^4$G$_{7/2}$   &   43083.05    &   2320.386    &   0.011   &   0.012   &   0.26    &   17  \\
        &   $3d^5$  $^4$G$_{5/2}$   &   43088.79    &   2320.077    &   0.101   &   0.118   &   2.29    &   15  \\
        &\textit{Residual}          &       &       &   0.009   &       &       &       \\
        &           &       &       &       &       &       &       \\
$3d^4$(a $^3$F)$4p$ $^4$F$_{3/2}^o$ &   $3d^4$($^1$S)$4s$   $^2$S$_{1/2}$   &   26655.36    &   3750.525    &   0.009   &   0.009   &   0.22    &   16  \\
        &   $3d^5$  $^2$F$_{5/2}$   &   27328.42    &   3658.160    &   0.031   &   0.005   &   0.79    &   12  \\
        &   $3d^4$($^3$F)$4s$   $^2$F$_{5/2}$   &   31501.25    &   3173.559    &   0.031   &   0.038   &   0.80    &   17  \\
        &   $3d^4$($^3$P)$4s$   $^2$P$_{3/2}$   &   31714.54    &   3152.215    &   0.493   &   0.464   &   12.63   &   8   \\
        &   $3d^4$($^3$P)$4s$   $^2$P$_{1/2}$   &   32411.20    &   3084.458    &   0.067   &   0.065   &   1.72    &   12  \\
        &   $3d^5$  $^2$D$_{5/2}$   &   35719.56    &   2798.762    &   0.105   &   0.136   &   2.70    &   11  \\
        &   $3d^4$($^3$F)$4s$   $^4$F$_{5/2}$   &   35953.09    &   2780.581    &   0.028   &   0.036   &   0.72    &   15  \\
        &   $3d^4$($^3$F)$4s$   $^4$F$_{3/2}$   &   35987.56    &   2777.920    &   0.027   &   0.051   &   0.69    &   14  \\
        &   $3d^4$($^3$P)$4s$   $^4$P$_{5/2}$   &   36205.99    &   2761.157    &   0.014   &   0.019   &   0.37    &   22  \\
        &   $3d^5$  $^4$P$_{1/2}$   &   45246.70    &   2209.417    &   0.021   &   0.019   &   0.55    &   18  \\
        &   $3d^5$  $^4$P$_{5/2}$   &   45247.91    &   2209.358    &   0.020   &   0.018   &   0.50    &   19  \\
        &   $3d^5$  $^4$G$_{5/2}$   &   46558.37    &   2147.165    &   0.068   &   0.054   &   1.73    &   13  \\
        &\textit{Residual}          &       &       &   0.087   &       &       &       \\
        &           &       &       &       &       &       &       \\
$3d^4$(a $^3$F)$4p$ $^4$F$_{5/2}^o$ &   $3d^4$($^1$D)$4s$   $^2$D$_{5/2}$   &   21281.50    &   4697.602    &   0.005   &   0.003   &   0.13    &   30  \\
        &   $3d^4$($^3$D)$4s$   $^2$D$_{5/2}$   &   24114.12    &   4145.780    &   0.029   &   0.022   &   0.77    &   12  \\
        &   $3d^5$  $^2$F$_{7/2}$   &   27135.04    &   3684.223    &   0.041   &   0.025   &   1.12    &   12  \\
        &   $3d^5$  $^2$F$_{5/2}$   &   27270.08    &   3665.979    &   0.062   &   0.002   &   1.69    &   12  \\
        &   $3d^4$($^1$G)$4s$   $^2$G$_{7/2}$   &   27328.36    &   3658.160    &   0.031   &   0.012   &   0.82    &   12  \\
        &   $3d^4$($^3$F)$4s$   $^2$F$_{7/2}$   &   31404.56    &   3183.330    &   0.042   &   0.221   &   1.13    &   12  \\
        &   $3d^4$($^3$F)$4s$   $^2$F$_{5/2}$   &   31442.88    &   3179.450    &   0.003   &   0.020   &   0.08    &   17  \\
        &   $3d^4$($^3$P)$4s$   $^2$P$_{3/2}$   &   31656.18    &   3158.027    &   0.101   &   0.061   &   2.73    &   12  \\
        &   $3d^4$($^3$G)$4s$   $^4$G$_{7/2}$   &   33491.04    &   2985.002    &   0.008   &   0.013   &   0.23    &   67  \\
        &   $3d^5$  $^2$F$_{5/2}$   &   34408.71    &   2905.390    &   0.018   &   0.013   &   0.47    &   39  \\
        &   $3d^5$  $^2$D$_{5/2}$   &   35661.19    &   2803.343    &   0.097   &   0.005   &   2.61    &   12  \\
        &   $3d^4$($^3$F)$4s$   $^4$F$_{7/2}$   &   35843.49    &   2789.084    &   0.029   &   0.043   &   0.77    &   14  \\
        &   $3d^4$($^3$F)$4s$   $^4$F$_{5/2}$   &   35894.74    &   2785.102    &   0.032   &   0.038   &   0.85    &   15  \\
        &   $3d^4$($^3$F)$4s$   $^4$F$_{3/2}$   &   35929.18    &   2782.432    &   0.027   &   0.034   &   0.73    &   15  \\
        &   $3d^4$($^3$P)$4s$   $^4$P$_{5/2}$   &   36147.63    &   2765.616    &   0.175   &   0.172   &   4.72    &   12  \\
        &   $3d^4$($^3$P)$4s$   $^4$P$_{3/2}$   &   36704.65    &   2723.644    &   0.024\footnote{Based on theoretical $A$-value from Ref.[1]} &   0.024   &   0.53    &   12  \\
        &   $3d^5$  $^4$D$_{7/2}$   &   41978.40    &   2381.451    &   0.010   &   0.012   &   0.26    &   59  \\
        &   $3d^5$  $^4$P$_{3/2}$   &   45187.88    &   2212.293    &   0.008   &   0.004   &   0.21    &   28  \\
        &   $3d^5$  $^4$P$_{5/2}$   &   45189.56    &   2212.211    &   0.034   &   0.020   &   0.92    &   18  \\
        &   $3d^5$  $^4$G$_{7/2}$   &   46494.27    &   2150.126    &   0.118   &   0.081   &   3.19    &   12  \\
        &   $3d^4$($^5$D)$4s$   $^6$D$_{7/2}$   &   46988.03    &   2127.530    &   0.042   &   0.021   &   1.13    &   22  \\
        &\textit{Residual}          &       &       &   0.072   &       &       &       \\
        &           &       &       &       &       &       &       \\
$3d^4$(a $^3$F)$4p$ $^4$F$_{7/2}^o$ &   $3d^4$($^3$G)$4s$   $^4$G$_{9/2}$   &   33774.50    &   2959.949    &   0.027   &   0.024   &   0.94    &   13  \\
        &   $3d^4$($^3$G)$4s$   $^4$G$_{7/2}$   &   33872.35    &   2951.398    &   0.020   &   0.018   &   0.69    &   16  \\
        &   $3d^5$  $^4$F$_{7/2}$   &   34556.79    &   2892.939    &   0.100   &   0.105   &   3.45    &   10  \\
        &   $3d^4$($^3$F)$4s$   $^4$F$_{9/2}$   &   36174.11    &   2763.592    &   0.047   &   0.073   &   1.62    &   10  \\
        &   $3d^4$($^3$F)$4s$   $^4$F$_{7/2}$   &   36224.86    &   2759.719    &   0.188   &   0.275   &   6.48    &   9   \\
        &   $3d^4$($^3$F)$4s$   $^4$F$_{5/2}$   &   36276.09    &   2755.822    &   0.016   &   0.028   &   0.56    &   13  \\
        &   $3d^4$($^3$P)$4s$   $^4$P$_{5/2}$   &   36528.98    &   2736.743    &   0.004   &   0.007   &   0.15    &   32  \\
        &   $3d^4$($^3$H)$4s$   $^4$H$_{9/2}$   &   37174.66    &   2689.206    &   0.093   &   0.127   &   3.20    &   10  \\
        &   $3d^4$($^3$H)$4s$   $^4$H$_{7/2}$   &   37236.70    &   2684.726    &   0.008   &   0.009   &   0.27    &   16  \\
        &   $3d^5$  $^4$D$_{5/2}$   &   42346.73    &   2360.736    &   0.005   &   0.007   &   0.16    &   22  \\
        &   $3d^5$  $^4$G$_{9/2}$   &   46874.16    &   2132.698    &   0.115   &   0.091   &   3.95    &   10  \\
        &   $3d^5$  $^4$G$_{7/2}$   &   46875.64    &   2132.631    &   0.274   &   0.165   &   9.45    &   8   \\
        &   $3d^4$($^5$D)$4s$   $^6$D$_{5/2}$   &   47595.59    &   2100.368    &   0.069   &   0.037   &   2.38    &   13  \\
        &\textit{Residual}          &       &       &   0.034   &       &       &       \\
        &           &       &       &       &       &       &       \\
$3d^4$(a $^3$F)$4p$ $^4$F$_{9/2}^o$ &   $3d^4$($^3$D)$4s$   $^4$D$_{7/2}$   &   29179.00    &   3426.140    &   0.002   &   0.003   &   0.07    &   22  \\
        &   $3d^4$($^3$H)$4s$   $^2$H$_{11/2}$  &   32635.58    &   3063.250    &   0.006   &   0.005   &   0.19    &   26  \\
        &   $3d^4$($^3$G)$4s$   $^4$G$_{9/2}$   &   33829.60    &   2955.127    &   0.010   &   0.011   &   0.36    &   24  \\
        &   $3d^5$  $^4$F$_{9/2}$   &   34594.28    &   2889.804    &   0.139   &   0.150   &   4.79    &   10  \\
        &   $3d^4$($^3$F)$4s$   $^4$F$_{9/2}$   &   36229.20    &   2759.389    &   0.292   &   0.419   &   10.07   &   9   \\
        &   $3d^4$($^3$F)$4s$   $^4$F$_{7/2}$   &   36279.96    &   2755.528    &   0.027   &   0.044   &   0.94    &   11  \\
        &   $3d^4$($^3$H)$4s$   $^4$H$_{11/2}$  &   37150.07    &   2690.986    &   0.029   &   0.032   &   1.00    &   11  \\
        &   $3d^5$  $^4$G$_{9/2}$   &   46929.24    &   2130.195    &   0.118   &   0.064   &   4.08    &   14  \\
        &   $3d^5$  $^4$G$_{11/2}$  &   46936.43    &   2129.869    &   0.271   &   0.187   &   9.35    &   9   \\
        &   $3d^4$($^5$D)$4s$   $^6$D$_{7/2}$   &   47424.51    &   2107.946    &   0.073   &   0.053   &   2.53    &   12  \\
        &\textit{Residual}          &       &       &   0.032   &       &       &       \\
        &           &       &       &       &       &       &       \\
$3d^4$(a $^3$H)$4p$ $^4$G$_{5/2}^o$ &   $3d^4$($^3$G)$4s$   $^4$G$_{5/2}$   &   33926.02    &   2946.729    &   0.036   &   0.038   &   1.39    &   31  \\
        &   $3d^5$  $^4$F$_{3/2}$   &   34499.28    &   2897.762    &   0.094   &   0.067   &   3.60    &   12  \\
        &   $3d^4$($^3$F)$4s$   $^4$F$_{3/2}$   &   36261.11    &   2756.960    &   0.061   &   0.044   &   2.34    &   11  \\
        &   $3d^4$($^3$H)$4s$   $^4$H$_{7/2}$   &   37187.26    &   2688.294    &   0.310   &   0.452   &   11.92   &   10  \\
        &   $3d^5$  $^4$G$_{7/2}$   &   46826.20    &   2134.883    &   0.093   &   0.045   &   3.57    &   12  \\
        &   $3d^5$  $^4$G$_{5/2}$   &   46831.93    &   2134.622    &   0.373   &   0.321   &   14.34   &   9   \\
        &\textit{Residual}          &       &       &   0.034   &       &       &       \\
        &           &       &       &       &       &       &       \\
$3d^4$(a $^3$H)$4p$ $^4$G$_{7/2}^o$ &   $3d^4$($^1$D)$4s$   $^2$D$_{5/2}$   &   21603.17    &   4627.654    &   0.005   &   0.000   &   0.21    &   23  \\
        &   $3d^4$($^3$G)$4s$   $^4$G$_{7/2}$   &   33812.68    &   2956.607    &   0.017   &   0.018   &   0.65    &   17  \\
        &   $3d^5$  $^4$F$_{5/2}$   &   34478.83    &   2899.480    &   0.084   &   0.083   &   3.22    &   12  \\
        &   $3d^5$  $^4$F$_{7/2}$   &   34497.16    &   2897.940    &   0.010   &   0.006   &   0.38    &   31  \\
        &   $3d^5$  $^2$F$_{5/2}$   &   34730.42    &   2878.476    &   0.025   &   0.013   &   0.94    &   28  \\
        &   $3d^5$  $^2$D$_{5/2}$   &   35982.90    &   2778.278    &   0.007   &   0.008   &   0.25    &   44  \\
        &   $3d^4$($^3$F)$4s$   $^4$F$_{9/2}$   &   36114.43    &   2768.159    &   0.014   &   0.014   &   0.55    &   15  \\
        &   $3d^4$($^3$F)$4s$   $^4$F$_{7/2}$   &   36165.19    &   2764.273    &   0.038   &   0.030   &   1.47    &   11  \\
        &   $3d^4$($^3$F)$4s$   $^4$F$_{5/2}$   &   36216.44    &   2760.361    &   0.090   &   0.114   &   3.46    &   10  \\
        &   $3d^4$($^3$H)$4s$   $^4$H$_{9/2}$   &   37114.99    &   2693.530    &   0.168   &   0.317   &   6.47    &   10  \\
        &   $3d^4$($^3$H)$4s$   $^4$H$_{7/2}$   &   37177.03    &   2689.035    &   0.021   &   0.036   &   0.80    &   12  \\
        &   $3d^5$  $^4$G$_{9/2}$   &   46814.49    &   2135.417    &   0.262   &   0.156   &   10.06   &   9   \\
        &   $3d^5$  $^4$G$_{7/2}$   &   46816.06    &   2135.345    &   0.228   &   0.179   &   8.76    &   9   \\
        &   $3d^5$  $^4$G$_{5/2}$   &   46821.68    &   2135.089    &   0.024   &   0.018   &   0.91    &   19  \\
        &\textit{Residual}          &       &       &   0.009   &       &       &       \\
        &           &       &       &       &       &       &       \\
$3d^4$(a $^3$H)$4p$ $^4$G$_{9/2}^o$ &   $3d^4$($^3$G)$4s$   $^4$G$_{9/2}$   &   33734.32    &   2963.474    &   0.019   &   0.019   &   0.71    &   22  \\
        &   $3d^5$  $^4$F$_{7/2}$   &   34516.61    &   2896.307    &   0.091   &   0.091   &   3.51    &   11  \\
        &   $3d^4$($^3$F)$4s$   $^4$F$_{7/2}$   &   36184.69    &   2762.784    &   0.075   &   0.094   &   2.87    &   11  \\
        &   $3d^4$($^3$H)$4s$   $^4$H$_{11/2}$  &   37054.78    &   2697.907    &   0.245   &   0.374   &   9.42    &   10  \\
        &   $3d^4$($^3$H)$4s$   $^4$H$_{9/2}$   &   37134.48    &   2692.116    &   0.026   &   0.047   &   1.00    &   11  \\
        &   $3d^5$  $^4$G$_{9/2}$   &   46833.98    &   2134.528    &   0.373   &   0.270   &   14.36   &   9   \\
        &   $3d^5$  $^4$G$_{7/2}$   &   46835.47    &   2134.460    &   0.036   &   0.025   &   1.38    &   13  \\
        &   $3d^5$  $^4$G$_{11/2}$  &   46841.16    &   2134.201    &   0.122   &   0.068   &   4.68    &   11  \\
        &\textit{Residual}          &       &       &   0.014   &       &       &       \\
        &           &       &       &       &       &       &       \\
$3d^4$(a $^3$H)$4p$ $^4$G$_{11/2}^o$    &   $3d^4$($^3$H)$4s$   $^2$H$_{9/2}$   &   32738.14    &   3053.653    &   0.009   &   0.004   &   0.34    &   19  \\
        &   $3d^4$($^3$G)$4s$   $^4$G$_{11/2}$  &   33674.95    &   2968.700    &   0.019   &   0.015   &   0.70    &   26  \\
        &   $3d^5$  $^4$F$_{9/2}$   &   34514.88    &   2896.452    &   0.096\footnote{Based on theoretical $A$-value from Ref.[1]} &   0.096   &   3.56    &   11  \\
        &   $3d^4$($^3$F)$4s$   $^4$F$_{9/2}$   &   36149.73    &   2765.456    &   0.071   &   0.097   &   2.64    &   12  \\
        &   $3d^4$($^3$H)$4s$   $^4$H$_{13/2}$  &   36977.23    &   2703.565    &   0.239   &   0.389   &   8.85    &   11  \\
        &   $3d^4$($^3$H)$4s$   $^4$H$_{11/2}$  &   37070.59    &   2696.756    &   0.020   &   0.032   &   0.75    &   12  \\
        &   $3d^5$  $^2$I$_{13/2}$  &   37219.20    &   2685.988    &   0.005   &   0.002   &   0.19    &   14  \\
        &   $3d^5$  $^4$G$_{9/2}$   &   46849.77    &   2133.809    &   0.026   &   0.016   &   0.96    &   20  \\
        &   $3d^5$  $^4$G$_{11/2}$  &   46856.96    &   2133.482    &   0.508   &   0.343   &   18.83   &   11  \\
        &\textit{Residual}          &       &       &   0.006   &       &       &       \\
        &           &       &       &       &       &       &       \\
$3d^4$(a $^3$F)$4p$ $^2$D$_{3/2}^o$ &   $3d^4$($^3$D)$4s$   $^2$D$_{3/2}$   &   24392.68    &   4098.434    &   0.006   &   0.006   &   0.19    &   27  \\
        &   $3d^5$  $^2$F$_{5/2}$   &   27637.24    &   3617.274    &   0.015   &   0.010   &   0.48    &   17  \\
        &   $3d^4$($^3$F)$4s$   $^2$F$_{5/2}$   &   31810.11    &   3142.745    &   0.050   &   0.049   &   1.62    &   13  \\
        &   $3d^4$($^3$P)$4s$   $^2$P$_{3/2}$   &   32023.39    &   3121.812    &   0.055   &   0.054   &   1.76    &   13  \\
        &   $3d^5$  $^4$F$_{5/2}$   &   34524.32    &   2895.660    &   0.021   &   0.025   &   0.69    &   21  \\
        &   $3d^5$  $^4$F$_{3/2}$   &   34534.58    &   2894.800    &   0.107   &   0.107   &   3.44    &   13  \\
        &   $3d^4$($^3$F)$4s$   $^4$F$_{5/2}$   &   36261.96    &   2756.896    &   0.062   &   0.083   &   1.99    &   13  \\
        &   $3d^4$($^3$F)$4s$   $^4$F$_{3/2}$   &   36296.41    &   2754.280    &   0.248   &   0.366   &   7.99    &   12  \\
        &   $3d^5$  $^4$G$_{5/2}$   &   46867.22    &   2133.015    &   0.337   &   0.215   &   10.87   &   11  \\
        &   $3d^4$($^5$D)$4s$   $^6$D$_{1/2}$   &   47851.04    &   2089.154    &   0.048   &   0.033   &   1.53    &   33  \\
        &\textit{Residual}          &       &       &   0.052   &       &       &       \\
        &           &       &       &       &       &       &       \\
$3d^4$(a $^3$F)$4p$ $^2$D$_{5/2}^o$ &   $3d^4$($^3$D)$4s$   $^2$D$_{5/2}$   &   24489.14    &   4082.291    &   0.003   &   0.006   &   0.10    &   32  \\
        &   $3d^5$  $^2$F$_{7/2}$   &   27510.06    &   3633.998    &   0.008   &   0.007   &   0.26    &   21  \\
        &   $3d^4$($^3$F)$4s$   $^2$F$_{7/2}$   &   31779.58    &   3145.763    &   0.042   &   0.055   &   1.35    &   13  \\
        &   $3d^4$($^3$P)$4s$   $^2$P$_{3/2}$   &   32031.19    &   3121.052    &   0.021   &   0.028   &   0.67    &   17  \\
        &   $3d^4$($^3$G)$4s$   $^4$G$_{7/2}$   &   33865.99    &   2951.953    &   0.022   &   0.019   &   0.71    &   20  \\
        &   $3d^5$  $^4$F$_{5/2}$   &   34532.14    &   2895.004    &   0.092   &   0.094   &   2.98    &   12  \\
        &   $3d^5$  $^2$F$_{5/2}$   &   34783.80    &   2874.058    &   0.029   &   0.008   &   0.92    &   17  \\
        &   $3d^4$($^3$F)$4s$   $^4$F$_{7/2}$   &   36218.51    &   2760.204    &   0.072   &   0.099   &   2.32    &   12  \\
        &   $3d^4$($^3$F)$4s$   $^4$F$_{5/2}$   &   36269.76    &   2756.303    &   0.199   &   0.298   &   6.42    &   11  \\
        &   $3d^4$($^3$F)$4s$   $^4$F$_{3/2}$   &   36304.20    &   2753.688    &   0.042   &   0.075   &   1.34    &   13  \\
        &   $3d^4$($^3$H)$4s$   $^4$H$_{7/2}$   &   37230.35    &   2685.183    &   0.017   &   0.008   &   0.55    &   15  \\
        &   $3d^5$  $^4$G$_{7/2}$   &   46869.29    &   2132.920    &   0.233   &   0.186   &   7.51    &   11  \\
        &   $3d^5$  $^4$G$_{5/2}$   &   46875.02    &   2132.659    &   0.109   &   0.041   &   3.50    &   13  \\
        &   $3d^4$($^5$D)$4s$   $^6$D$_{3/2}$   &   47755.83    &   2093.320    &   0.044   &   0.010   &   1.42    &   21  \\
        &\textit{Residual}          &       &       &   0.068   &       &       &       \\

\hline

\end{longtable}
\noindent $^a$\citeauthor{RU} (\citeyear{RU})\\
$^b$Based on theoretical $A$-value from \citeauthor{RU} (\citeyear{RU})

}



The $a$$^4$P$_{3/2}$-$y$$^4$F$_{5/2}$ transition at 36704.65\,cm$^{-1}$ and the $a$$^4$F$_{9/2}$-$y$$^4$G$_{11/2}$ transition at 34514.88\,cm$^{-1}$ were found
to be considerably stronger than predicted by semi-empirical calculations. This deviation stems from line blending with other \ion{Cr}{ii} transitions. The
line at 36704.65\,cm$^{-1}$ is within 0.05\,cm$^{-1}$ of a line of comparable strength from the $b$\,$^4$G$_{11/2}$-$y$\,$^2$H$_{11/2}$ transition and the line
at 34514.88\,cm$^{-1}$ is within 0.05\,cm$^{-1}$ of the $b$\,$^2$I$_{11/2}$-$z$$^2$I$_{13/2}$ transition. We observed other relatively strong transitions from
these $y$\,$^2$H$_{11/2}$ and $z$$^2$I$_{13/2}$ upper levels, which indicated that these levels had relatively high populations and were thus probable
candidates for line blending. However, it was impossible to fit the two blended features to recover the individual line intensities because of the close
proximity of the central wavenumbers of the transitions. Instead we used $BF$s  from the semi-empirical calculations of \citeauthor{RU} (\citeyear{RU}) for the
36704.65\,cm$^{-1}$ and 34514.88\,cm$^{-1}$ transitions.



\onllongtab{3}{
\begin{longtable}{@{}cccccc}
\caption{\label{Linelist} Line list and experimental $\log{(gf)}$-values together with calculated values from the literature.}\\
\hline\hline
$\lambda_{vac}$ & $\sigma$ & $E_{lower}$ & \multicolumn{3}{c}{log~$gf$}\\
\cline{4-6}
(\AA) & (cm$^{-1})$ & (cm$^{-1})$ &  This study  &  K$^a$ & R\&U$^b$\\
\hline
\endfirsthead
\caption{continued.}\\
\hline\hline
$\lambda_{vac}$ & $\sigma$ & $E_{lower}$ & \multicolumn{3}{c}{log~$gf$}\\
\cline{4-6}
(\AA) & (cm$^{-1})$ & (cm$^{-1})$ &  This study  & K$^a$ & R\&U$^b$\\
\hline
\endhead
\hline
\endfoot

2089.819    &   47851.04    &   11961.81    &  $-$1.396  &  $-$1.649  &  $-$1.59   \\
2093.985    &   47755.83    &   12032.58    &  $-$1.252  &  $-$1.527  &  $-$1.95   \\
2101.035    &   47595.59    &   12147.82    &  $-$0.899  &  $-$1.078  &  $-$1.16   \\
2108.614    &   47424.51    &   12303.86    &  $-$0.746  &  $-$0.906  &  $-$0.91   \\
2128.202    &   46988.03    &   12303.86    &  $-$1.316  &  $-$2.172  &  $-$1.70   \\
2129.869    &   46936.43    &   20512.10    &  $-$0.169  &  $-$0.091  &  $-$0.36   \\
2130.195    &   46929.24    &   20519.33    &  $-$2.277  &  $-$0.556  &  $-$0.82   \\
2133.304    &   46875.64    &   20517.83    &  $-$0.288  &  $-$2.465  &  $-$0.50   \\
2133.332    &   46875.02    &   20512.06    &  $-$0.843  &  $-$2.234  &  $-$1.31   \\
2133.372    &   46874.16    &   20519.33    &  $-$0.666  &  $-$0.201  &  $-$0.76   \\
2133.593    &   46869.29    &   20517.83    &  $-$0.512  &  $-$0.577  &  $-$0.64   \\
2133.687    &   46867.22    &   20512.06    &  $-$0.528  &  $-$0.682  &  $-$0.76   \\
2134.155    &   46856.96    &   20512.10    &  $+$0.188  &  $+$0.190  &  $+$0.05   \\
2134.482    &   46849.77    &   20519.33    &  $-$1.105  &  $-$1.097  &  $-$1.24   \\
2134.875    &   46841.16    &   20512.10    &  $-$0.495  &  $-$1.410  &  $-$0.73   \\
2135.134    &   46835.47    &   20517.83    &  $-$1.027  &  $-$0.942  &  $-$1.17   \\
2135.202    &   46833.98    &   20519.33    &  $-$0.008  &  $+$0.094  &  $-$0.13   \\
2135.295    &   46831.93    &   20512.06    &  $-$0.231  &  $-$0.091  &  $-$0.28   \\
2135.557    &   46826.20    &   20517.83    &  $-$0.835  &  $-$1.268  &  $-$1.14   \\
2135.763    &   46821.68    &   20512.06    &  $-$1.303  &  $-$1.047  &  $-$1.41   \\
2136.019    &   46816.06    &   20517.83    &  $-$0.319  &  $-$0.005  &  $-$0.42   \\
2136.091    &   46814.49    &   20519.33    &  $-$0.259  &  $-$1.329  &  $-$0.48   \\
2147.841    &   46558.37    &   20512.06    &  $-$1.319  &  $-$1.010  &  $-$1.48   \\
2150.126    &   46494.28    &   20517.83    &  $-$0.713  &  $-$0.877  &  $-$1.10   \\
2210.047    &   45247.91    &   21822.52    &  $-$1.833  &  $-$1.890  &  $-$1.93   \\
2210.106    &   45246.70    &   21823.84    &  $-$1.795  &  $-$2.353  &  $-$1.92   \\
2212.900    &   45189.56    &   21822.52    &  $-$1.371  &  $-$1.840  &  $-$1.68   \\
2212.983    &   45187.88    &   21824.11    &  $-$2.021  &  $-$3.826  &  $-$2.34   \\
2297.879    &   43518.40    &   20512.10    &  $-$0.483  &  $-$0.341  &  $-$0.46   \\
2307.519    &   43336.59    &   20512.10    &  $-$1.694  &  $-$1.697  &  $-$1.70   \\
2307.901    &   43329.42    &   20519.33    &  $-$0.594  &  $-$0.443  &  $-$0.58   \\
2315.434    &   43188.45    &   20517.83    &  $-$0.736  &  $-$0.540  &  $-$0.68   \\
2315.513    &   43186.97    &   20519.33    &  $-$1.630  &  $-$1.589  &  $-$1.62   \\
2320.789    &   43088.79    &   20512.06    &  $-$0.830  &  $-$0.631  &  $-$0.76   \\
2321.099    &   43083.05    &   20517.83    &  $-$1.777  &  $-$1.709  &  $-$1.76   \\
2382.178    &   41978.40    &   25033.70    &  $-$1.853  &  $-$2.462  &  $-$1.83   \\
2685.523    &   37236.70    &   30156.79    &  $-$1.627  &  $-$2.209  &  $-$1.57   \\
2685.981    &   37230.35    &   30156.79    &  $-$1.450  &  $-$2.133  &  $-$1.80   \\
2686.785    &   37219.20    &   30149.83    &  $-$1.612  &  $-$1.588  &  $-$2.02   \\
2689.093    &   37187.26    &   30156.79    &  $-$0.111  &  $+$0.115  &  $+$0.07   \\
2689.833    &   37177.03    &   30156.79    &  $-$1.159  &  $-$0.841  &  $-$0.91   \\
2690.004    &   37174.66    &   30218.81    &  $-$0.556  &  $-$1.473  &  $-$0.41   \\
2691.785    &   37150.07    &   30298.51    &  $-$0.936  &  $-$1.726  &  $-$0.92   \\
2692.915    &   37134.48    &   30218.81    &  $-$0.962  &  $-$0.701  &  $-$0.69   \\
2694.329    &   37114.99    &   30218.81    &  $-$0.249  &  $+$0.193  &  $+$0.03   \\
2697.556    &   37070.59    &   30298.51    &  $-$1.005  &  $-$0.795  &  $-$0.79   \\
2698.707    &   37054.78    &   30298.51    &  $+$0.012  &  $+$0.276  &  $+$0.21   \\
2704.367    &   36977.23    &   30391.83    &  $+$0.066  &  $+$0.344  &  $+$0.31   \\
2724.450    &   36704.65    &   30307.44    &  $-$1.451\footnote{Unresolved blend with the line $70398.87_{11/2}$-$33694.15_{11/2}$}    &  $-$2.952  &  $-$1.42 \\
2737.552    &   36528.98    &   30864.46    &  $-$1.872  &  $-$1.572  &  $-$1.69   \\
2754.502    &   36304.20    &   31082.94    &  $-$1.038  &  $-$0.787  &  $-$0.82   \\
2755.093    &   36296.41    &   31082.94    &  $-$0.439  &  $-$0.310  &  $-$0.30   \\
2756.343    &   36279.96    &   31168.58    &  $-$0.945  &  $-$0.382  &  $-$0.76   \\
2756.637    &   36276.09    &   31117.39    &  $-$1.289  &  $-$0.412  &  $-$1.05   \\
2757.118    &   36269.76    &   31117.39    &  $-$0.358  &  $-$0.295  &  $-$0.22   \\
2757.711    &   36261.96    &   31117.39    &  $-$1.042  &  $-$0.859  &  $-$0.94   \\
2757.775    &   36261.11    &   31082.94    &  $-$0.796  &  $-$1.168  &  $-$0.92   \\
2760.204    &   36229.20    &   31219.35    &  $+$0.088  &  $+$0.260  &  $+$0.22   \\
2760.534    &   36224.87    &   31168.58    &  $-$0.228  &  $-$0.015  &  $-$0.05   \\
2761.019    &   36218.51    &   31168.58    &  $-$0.798  &  $-$0.765  &  $-$0.69   \\
2761.177    &   36216.44    &   31117.39    &  $-$0.499  &  $-$1.141  &  $-$0.39   \\
2761.973    &   36205.00    &   30864.46    &  $-$1.775  &  $-$4.440  &  $-$1.72   \\
2763.600    &   36184.69    &   31168.58    &  $-$0.483  &  $-$0.710  &  $-$0.37   \\
2764.408    &   36174.11    &   31219.35    &  $-$0.829  &  $-$0.414  &  $-$0.63   \\
2765.090    &   36165.19    &   31168.58    &  $-$0.870  &  $-$0.951  &  $-$0.97   \\
2766.272    &   36149.73    &   31219.35    &  $-$0.440  &  $-$0.294  &  $-$0.28   \\
2766.433    &   36147.63    &   30864.46    &  $-$0.468  &  $-$0.997  &  $-$0.55   \\
2768.976    &   36114.43    &   31219.35    &  $-$1.299  &  $-$1.876  &  $-$1.73   \\
2778.740    &   35987.53    &   31082.94    &  $-$1.494  &  $-$0.886  &  $-$1.28   \\
2779.098    &   35982.90    &   31350.90    &  $-$1.634  &  $-$1.635  &  $-$1.54   \\
2781.401    &   35953.10    &   31117.39    &  $-$1.476  &  $-$1.280  &  $-$1.43   \\
2783.253    &   35929.18    &   31082.94    &  $-$1.274  &  $-$1.010  &  $-$1.25   \\
2785.923    &   35894.74    &   31117.39    &  $-$1.204  &  $-$0.885  &  $-$1.20   \\
2789.907    &   35843.49    &   31168.58    &  $-$1.246  &  $-$0.670  &  $-$1.15   \\
2799.587    &   35719.56    &   31350.90    &  $-$0.897  &  $-$0.877  &  $-$0.85   \\
2804.169    &   35661.19    &   31350.90    &  $-$0.714  &  $-$0.581  &  $-$0.85   \\
2874.902    &   34783.80    &   32603.40    &  $-$1.164  &  $-$2.739  &  $-$1.73   \\
2879.320    &   34730.42    &   32603.40    &  $-$1.027  &  $-$2.233  &  $-$1.30   \\
2890.651    &   34594.28    &   32854.31    &  $-$0.195  &  $-$0.302  &  $-$0.19   \\
2893.787    &   34556.79    &   32836.68    &  $-$0.461  &  $-$0.585  &  $-$0.43   \\
2895.648    &   34534.58    &   32844.76    &  $-$0.762  &  $-$0.920  &  $-$0.79   \\
2895.853    &   34532.14    &   32854.95    &  $-$0.649  &  $-$0.895  &  $-$0.68   \\
2896.509    &   34524.32    &   32854.95    &  $-$1.462  &  $-$1.546  &  $-$1.42   \\
2897.156    &   34516.61    &   32836.68    &  $-$0.355  &  $-$0.427  &  $-$0.34   \\
2897.301    &   34514.88    &   32854.31    &  $-$0.270\footnote{Unresolved blend with the line $74743.28_{13/2}$-$40228.33_{11/2}$}    &  $-$0.219  &  $-$0.24\\
2898.611    &   34499.28    &   32844.76    &  $-$0.565  &  $-$0.724  &  $-$0.70   \\
2898.789    &   34497.16    &   32836.68    &  $-$1.417  &  $-$1.112  &  $-$1.62   \\
2900.330    &   34478.83    &   32854.95    &  $-$0.489  &  $-$0.623  &  $-$0.49   \\
2906.241    &   34408.71    &   32603.40    &  $-$1.423  &  $-$1.594  &  $-$1.62   \\
2947.590    &   33926.02    &   33417.99    &  $-$0.960  &  $-$0.841  &  $-$0.92   \\
2951.541    &   33880.61    &   30149.83    &  $-$1.531  &  $-$2.319  &  $-$2.59   \\
2952.260    &   33872.35    &   33521.11    &  $-$1.139  &  $-$3.016  &  $-$1.18   \\
2952.815    &   33865.99    &   33521.11    &  $-$1.252  &  $-$2.163  &  $-$1.36   \\
2955.992    &   33829.60    &   33618.94    &  $-$1.302  &  $-$2.009  &  $-$1.29   \\
2957.470    &   33812.68    &   33521.11    &  $-$1.164  &  $-$0.790  &  $-$1.14   \\
2960.814    &   33774.50    &   33618.94    &  $-$1.004  &  $-$1.535  &  $-$1.05   \\
2964.340    &   33734.32    &   33618.94    &  $-$1.028  &  $-$0.776  &  $-$1.01   \\
2969.566    &   33674.95    &   33694.15    &  $-$0.958  &  $-$0.918  &  $-$1.02   \\
2972.769    &   33638.67    &   30391.83    &  $+$0.558  &  $+$0.559  &  $+$0.52   \\
2980.606    &   33550.22    &   30298.51    &  $+$0.472  &  $+$0.465  &  $+$0.43   \\
2985.873    &   33491.04    &   33521.11    &  $-$1.722  &  $-$1.625  &  $-$1.63   \\
2986.192    &   33487.47    &   30218.81    &  $+$0.364  &  $+$0.368  &  $+$0.33   \\
2988.923    &   33456.86    &   30391.83    &  $-$0.840  &  $-$0.800  &  $-$0.90   \\
2990.062    &   33444.12    &   30156.79    &  $+$0.271  &  $+$0.273  &  $+$0.23   \\
2993.315    &   33407.78    &   30298.51    &  $-$0.760  &  $-$0.690  &  $-$0.78   \\
2995.620    &   33382.07    &   30218.81    &  $-$0.864  &  $-$0.825  &  $-$0.91   \\
3054.541    &   32738.14    &   34630.95    &  $-$1.240  &  $-$2.370  &  $-$1.54   \\
3064.422    &   32635.58    &   34812.95    &  $-$1.536  &  $-$1.584  &  $-$1.65   \\
3064.725    &   32629.36    &   31219.35    &  $-$1.337  &  $-$1.737  &  $-$1.50   \\
3073.361    &   32537.67    &   31168.58    &  $-$1.393  &  $-$1.756  &  $-$1.51   \\
3078.483    &   32483.54    &   31117.39    &  $-$1.633  &  $-$2.185  &  $-$1.86   \\
3085.354    &   32411.20    &   34659.32    &  $-$1.009  &  $-$2.756  &  $-$1.09   \\
3121.957    &   32031.19    &   35355.89    &  $-$1.234  &  $-$0.827  &  $-$1.32   \\
3122.717    &   32023.39    &   35355.89    &  $-$0.988  &  $-$1.198  &  $-$1.02   \\
3143.655    &   31810.11    &   35569.20    &  $-$1.017  &  $-$0.789  &  $-$1.06   \\
3146.675    &   31779.58    &   35607.50    &  $-$0.919  &  $-$0.471  &  $-$0.84   \\
3153.128    &   31714.54    &   35355.89    &  $-$0.123  &  $-$0.559  &  $-$0.21   \\
3158.941    &   31656.18    &   35355.89    &  $-$0.590  &  $-$0.956  &  $-$0.89   \\
3174.477    &   31501.25    &   35569.20    &  $-$1.317  &  $-$0.561  &  $-$1.29   \\
3180.370    &   31442.88    &   35569.20    &  $-$2.119  &  $-$1.357  &  $-$2.48   \\
3184.251    &   31404.56    &   35607.50    &  $-$0.965  &  $-$0.380  &  $-$0.32   \\
3226.387    &   30994.42    &   32854.31    &  $-$2.126  &  $-$1.933  &  $-$1.75   \\
3239.439    &   30869.55    &   32836.68    &  $-$1.966  &  $-$1.942  &  $-$1.76   \\
3296.375    &   30336.36    &   33694.15    &  $-$0.648  &  $-$0.285  &  $-$0.33   \\
3307.999    &   30229.75    &   33618.94    &  $-$0.712  &  $-$0.368  &  $-$0.43   \\
3312.887    &   30185.16    &   33521.11    &  $-$0.808  &  $-$0.447  &  $-$0.51   \\
3313.137    &   30182.87    &   33417.99    &  $-$0.874  &  $-$0.523  &  $-$0.58   \\
3316.248    &   30154.56    &   33694.15    &  $-$1.984  &  $-$1.695  &  $-$1.67   \\
3323.660    &   30087.32    &   33618.94    &  $-$1.821  &  $-$1.542  &  $-$1.54   \\
3324.495    &   30079.75    &   33521.11    &  $-$1.988  &  $-$1.624  &  $-$1.63   \\
3422.572    &   29217.80    &   34630.95    &  $-$2.081  &  $-$2.041  &  $-$2.22   \\
3422.595    &   29217.60    &   34812.95    &  $-$1.863  &  $-$1.467  &  $-$1.65   \\
3427.122    &   29179.00    &   38269.59    &  $-$1.856  &  $-$1.658  &  $-$1.71   \\
3618.306    &   27637.24    &   32603.40    &  $-$1.423  &  $-$1.288  &  $-$1.62   \\
3635.034    &   27510.06    &   32355.68    &  $-$1.508  &  $-$1.008  &  $-$1.64   \\
3659.202    &   27328.36    &   36272.54    &  $-$1.199$^c$  &  $-$4.526  &  $-$4.54   \\
3659.202    &   27328.36    &   32603.40    &  $-$1.199$^c$  &  $-$1.142  &  $-$2.04   \\
3659.202    &   27328.36    &   39683.75    &  $-$1.199$^c$  &  $-$2.305  &  $-$1.45   \\
3667.023    &   27270.08    &   32603.40    &  $-$0.670  &  $-$2.327  &  $-$2.37   \\
3685.271    &   27135.04    &   32355.68    &  $-$0.844  &  $-$0.946  &  $-$1.15   \\
3751.591    &   26655.36    &   40415.09    &  $-$1.724  &  $-$2.537  &  $-$1.77   \\
4083.443    &   24489.14    &   42897.99    &  $-$1.816  &  $-$1.234  &  $-$1.58   \\
4099.591    &   24392.68    &   42986.62    &  $-$1.713  &  $-$1.470  &  $-$1.75   \\
4146.948    &   24114.12    &   42897.99    &  $-$0.902  &  $-$1.164  &  $-$1.11   \\
4628.950    &   21603.17    &   45730.58    &  $-$1.269  &  $-$4.682  &  $-$4.74   \\
4698.916    &   21281.50    &   45730.58    &  $-$1.584  &  $-$1.619  &  $-$1.91   \\

\hline

\end{longtable}
\noindent$^a$ Kurucz (1988)\\
$^b$ \citeauthor{RU} (\citeyear{RU})\\
$^c$ Use of the theoretical values are encouraged due to line blending.\\
}



\begin{figure}
\begin{center}
\includegraphics[width=8cm]{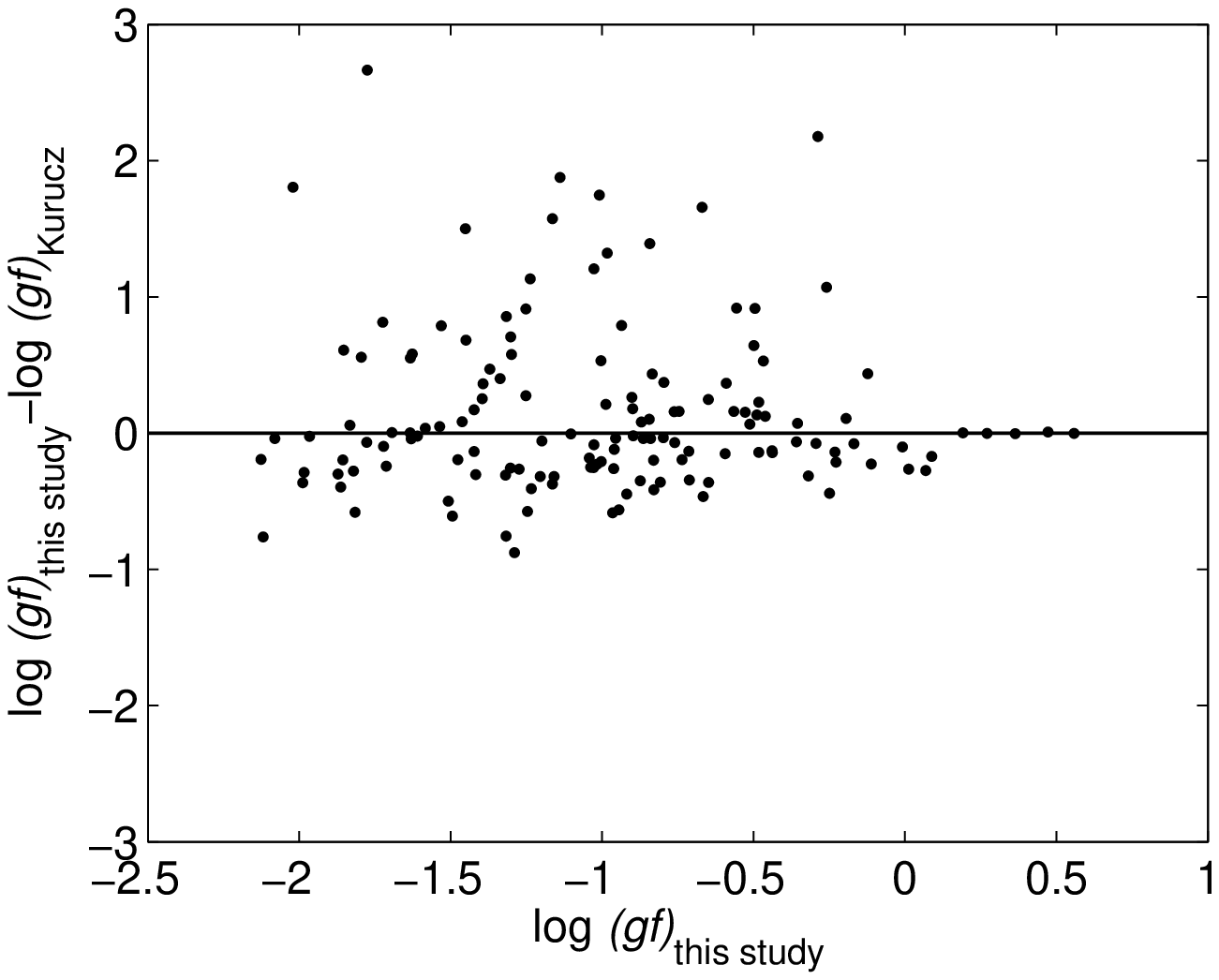}
\caption{Comparison between $\log{(gf)}$-values from this experimental study and semi-empirical values from \citeauthor{Kurucz2}~(\citeyear{Kurucz2}). The difference between the two values for each line is plotted as a function of line strength.} \label{loggf}
\end{center}
\end{figure}

\begin{figure}
\begin{center}
\includegraphics[width=8cm]{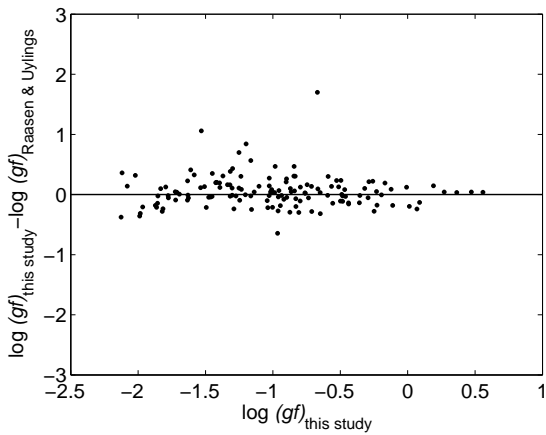}
\caption{Comparison between $\log{(gf)}$-values from this experimental study and semi-empirical values from \citeauthor{RU} (\citeyear{RU}). The difference
between the two values for each line is plotted as a function of line strength.} \label{RU}
\end{center}
\end{figure}

The log~$gf$s in Table~\ref{Linelist} are compared to the semi-empirical results by \citeauthor{Kurucz2}~(\citeyear{Kurucz2}) and
\citeauthor{RU}~(\citeyear{RU}) and a graphical comparison is given in Figs.~\ref{loggf} and \ref{RU}. The comparison between our laboratory log~$gf$s and
those of \citeauthor{RU} (\citeyear{RU}), Figure~\ref{RU}, has a  smaller deviation than those in Figure~\ref{loggf}.

There is good agreement between our experimental log~$gf$ values and the semi-empirical log~$gf$ values of \citeauthor{Kurucz2} (\citeyear{Kurucz2}) for lines
from the $z$\,$^4$H$_J$ levels as well as for lines from the $y$\,$^4$G$_{5/2}$ and the $z$\,$^2$D$_{3/2}$ levels, see Table~\ref{Linelist}. However, there is
a larger deviation between our log~$gf$ values and the semi-empirical calculations of \citeauthor{Kurucz2} (\citeyear{Kurucz2}) for lines from the
$y$\,$^4$F$_J$ levels, the $y$\,$^4$G$_{7/2, 9/2, 11/2}$ levels, and the $z$\,$^2$D$_{5/2}$ level. This may be due to inaccurate level mixing in the
semi-empirical calculations of \citeauthor{Kurucz2} (\citeyear{Kurucz2}). The semi-empirical calculations of \citeauthor{RU} (\citeyear{RU}) agree more
consistently with our laboratory log~$gf$ values.

In Table~\ref{Acomp} we compare our $A$-values with $A$-values in the literature. In general our $A$-values agree to within the uncertainties with the results
in the literature; however, the $A$-values of \citeauthor{Corliss}~\citeyearpar{Corliss} are significantly different to the other $A$-values in the literature.
The compilation of \citeauthor{Corliss}~\citeyearpar{Corliss} includes many values determined from a wall-stabilized arc. Inaccuracies in the wall temperature
measurement for the wall-stabilized arc method can significantly increase the uncertainty in the measurement, so we recommend that the $A$-values in
\citeauthor{Corliss}~\citeyearpar{Corliss} are used with caution.

\begin{table*}
 \caption{A comparison between our $A$-values and the $A$-values in the literature.} \label{Acomp} \centering
\resizebox{\textwidth}{!}{%
\begin{tabular}{@{}@{}llccccccccccc}
  \hline\hline
Upper & Lower &  $\sigma$ & \multicolumn{9}{c}{$A (10^8$\,s$^{-1}$)}\\
\cline{4-12}
level & level &  (cm$^{-1}$)  & This study & R\&U$^a$& K$^b$ & W\&W$^c$ & G\&W$^d$ & M$^e$ & Wt$^f$ & We$^g$ & C\&B$^h$\\
\hline
  $3d^4$($^3$H)$4p$ $^4$H$_{13/2}^o$  &   $3d^4$($^3$G)$4s$ $^4$G$_{11/2}$    &   30336.36    & 0.099 & 0.207 & 0.318 &   &  & 0.33  &  0.17 & 0.17$^i$ & 0.86\\
  $3d^4$($^3$H)$4p$ $^4$H$_{13/2}^o$  &   $3d^4$($^3$H)$4s$ $^4$H$_{13/2}$    &   33638.67    & 1.951 & 1.787 & 2.003 &   &   & 2.0 &  1.5 & 1.5$^i$ & 5.1\\
  $3d^4$($^3$H)$4p$ $^4$H$_{11/2}^o$  &   $3d^4$($^3$H)$4s$ $^4$H$_{11/2}$    &   33550.22    & 1.851 & 1.672 & 1.804  &   &   & 1.9 &  1.6 & 1.7$^i$ & 5.1\\
  $3d^4$($^3$H)$4p$ $^4$H$_{9/2}^o$&   $3d^4$($^3$H)$4s$ $^4$H$_{9/2}$ &   33487.47    & 1.720 & 1.600 & 2.207 &   &   & 2.3 &  1.7 & 1.8$^i$ & 3.6\\
  $3d^4$($^3$H)$4p$ $^4$H$_{7/2}^o$&   $3d^4$($^3$H)$4s$ $^4$H$_{7/2}$ &   33444.12    & 1.741 & 1.602 & 2.186 &   &  & 2.3 &  0.52 & 0.55$^i$ & 6.1\\
  $3d^4$($^3$H)$4p$ $^4$G$_{9/2}^o$&   $3d^4$($^3$H)$4s$ $^4$H$_{9/2}$ &   37134.48    & 0.100 & 0.186 & 0.183 &   & 0.13(7)  &  &  &  &\\
  $3d^4$($^3$F)$4p$ $^4$F$_{5/2}^o$&   $3d^4$($^1$D)$4s$ $^2$D$_{5/2}$ &   21281.50    & 0.011 & 0.006 & 0.007 & 0.046(9)  &   &  & 0.0492 &  & \\
  \hline

 \multicolumn{6}{l}{$^a$\citeauthor{RU} (\citeyear{RU})}\\
 \multicolumn{6}{l}{$^b$Kurucz (1988)}\\
 \multicolumn{6}{l}{$^c$Wujec and Weniger (1981)}\\
 \multicolumn{6}{l}{$^d$Goly and Weniger (1980) }\\
 \multicolumn{6}{l}{$^e$Musielok (1975) }\\
 \multicolumn{6}{l}{$^f$Warner (1967) (Calculated)}\\
 \multicolumn{6}{l}{$^g$Warner (1967) (Experimental)}\\
 \multicolumn{6}{l}{$^h$Corliss and Bozman (1962)}\\
\multicolumn{7}{l}{$^i$Upper limit}\\
\end{tabular}}
\end{table*}

\section{Summary}

We present experimental lifetimes for 14 highly excited energy levels in \ion{Cr}{ii} and $BF$s for 145 transitions from these levels, yielding experimental
transition probabilities for lines that are strong features in stellar spectra. For the majority of the \ion{Cr}{ii} transitions in this paper, the
experimental transition probabilities agree with the semi-empirical calculations within the uncertainty of the measurements. In particular, we note that the
semi-empirical orthogonal operator calculations of \citeauthor{RU}~(\citeyear{RU}) have a one standard deviation difference to our experimental log~$gf$ values
of 0.21. However, the semi-empirical Cowan code calculations of \citeauthor{Kurucz2} (\citeyear{Kurucz2}) have a one standard deviation difference to our
experimental log~$gf$ values of 0.61. In addition, different semi-empirical calculations produce different term labels for the energy levels because of the
large amount of level mixing. We suggested that further theoretical calculations for \ion{Cr}{ii} would benefit studies of this ion.

\begin{acknowledgements}
JG and SM are very grateful for the warm hospitality shown by the staff at Lund Observatory. Financial support from the Swedish Research Council (VR), a
Linnaeus grant to the Lund Laser Centre and the Knut and Alice Wallenberg Foundation, is gratefully acknowledged. RBW acknowledges a Euopean Commision Marie
Curie Intra-European fellowship.
\end{acknowledgements}

\end{document}